\documentclass[journal]{IEEEtran}
\usepackage{cite}
\usepackage[pdftex]{graphicx}
\usepackage{amsmath}
\usepackage{amssymb}
\usepackage{array}
\usepackage{csquotes}
\usepackage[numbers,sort&compress]{natbib}
\usepackage{color,soul}
\begin{document}
\title{Equalization Methods for NLIN Mitigation}
%
%
%

\author{Ori~Golani,
	Meir~Feder,
	and~Mark~Shtaif%
\thanks{O. Golani, M. Feder, and M. Shtaif are with the School of Electrical Engineering, Tel Aviv University, Tel Aviv, Israel 69978.
}
\thanks{Manuscript received January ?, 2017; revised May ?, 2017.}}

\markboth{Journal of Lightwave Technology,~Vol.~??, No.~?, May~2017}%
{name of paper here}

\maketitle

\begin{abstract}
We investigate the potential of adaptive equalization techniques to mitigate
inter-channel nonlinear interference noise (NLIN).
We derive a lower bound on the mutual information of a system using adaptive equalization, showing that the channel estimation error determines the equalizer's performance.
We develop an adaptive equalization scheme which uses the statistics of the NLIN to obtain optimal detection, based on Kalman filtering and maximum likelihood sequence estimation (MLSE). This scheme outperforms commonly used equalizers and significantly increases performance. 
\end{abstract}


%
\IEEEpeerreviewmaketitle

\section{Introduction}
%
%
%
%
\IEEEPARstart{I}n recent years, the challenge of mitigating nonlinear interference in wavelength division multiplexed (WDM) networks has attracted significant efforts. It has been well established that in typical WDM systems, where the number of channels is characteristically large and where the symbol rate is well over 10 Gigabaud, the nonlinear interference is dominated by the cross-phase-modulation (XPM) phenomenon, which has been shown to manifest itself predominantly in the form of linear, time-varying, intersymbol interference (ISI) \cite{secondini2012analytical,dar2015inter}. This reality calls for the adoption of ISI equalization methods for the mitigation of nonlinear penalties in fiber-optic WDM systems \cite{secondini2014xpm}.

The unique feature of the ISI induced by the nonlinear interference in fibers is that the ISI coefficients are dependent on the data transmitted in the multiple interfering channels (ICs), to which the receiver has no access. Therefore, these ISI coefficients are unknown and need to be estimated from the received signal. The main challenge facing the equalization procedure is that the ISI coefficients may change fairly fast, with the exact correlation times depending on the various system parameters \cite{secondini2012analytical,golani2016modeling,golani2016correlations}. The best equalization so far was demonstrated using the relatively fast converging recursive least squares (RLS) algorithm, which produced SNR improvements of the order of 0.5dB in a 16 QAM 80 channels system implemented over $5\times100$ km of SMF \cite{golani2016modeling}, whereas notably smaller benefits were observed in longer WDM links. This, as well as all other published  work, concentrated on the equalization of the zeroth-order ISI term (known as the phase and polarization rotation noise -- PPRN ) \cite{dar2016limits,li2010nonlinear,yankov2015low,zibar2016machine}, whose evolution is characterized by the longest correlation time \cite{dar2015inter}. Indeed, excluding the unrealistic case of ideally distributed amplification, higher order ISI terms have been resistant to equalization attempts \cite{secondini2014xpm}, as they change too rapidly for conventional equalization algorithms to track \cite{golani2016correlations,dar2015inter}. 
Today, in spite of the recent advances in understanding the impact of nonlinear propagation penalties, the prospect of NLIN mitigation by means of equalization remains largely unclear.
 
In this work we explore the potential limits of equalization as a tool for mitigating the effects of NLIN.
In order to do this, we capitalize on the detailed statistical model of NLIN that was derived in \cite{golani2016correlations,golani2016modeling}, and develop an adaptive algorithm that is customized for tracking the rapid variations of the high order ISI coefficients. 
This is in contrast to generic equalization algorithms that were used in the past, and which were oblivious to the actual statistical properties of the NLIN.  
By using information theoretic bounds on the capacity of partially known ISI channels, we obtain a lower bound for the achievable information rate (AIR) in a WDM system, which turns out to be notably higher than that obtained by treating the NLIN simply as additive Gaussian noise (AGN). 
Finally, we develop an approximated maximum likelihood sequence estimation (MLSE) equalizer for improving the raw BER of the channel, prior to forward error correction (FEC). 

It is important to note that the method that we describe is very intensive computationally, which makes its implementation very challenging in the context of  real-time WDM systems. Hence, we view the main importance of this work in demonstrating the potential of NLIN mitigation using equalization, and in demonstrating the benefit of using the correct statistical model. We demonstrate  gains in excess of 3dB in the effective SNR (ESNR) of the channel, which translate into a correspondingly large enhancement in the AIR. When considering the pre-FEC BER, gains of up to 0.8dB in terms of the system's Q-factor are observed.

The remainder of this paper is organized as follows; After introducing the notation in section \ref{notation}, we discuss in Sec. \ref{Achievable_information_section} the relevant aspects of NLIN, and consider the implications of this model to channel capacity. A lower bound on the mutual information that can be obtained using equalization, is derived in section \ref{MI_section}, while Sec. \ref{kalman_section} is dedicated to designing an algorithm for estimating the ISI coefficients by means of a Kalman filter. Section \ref{MLSE_section} combines the Kalman filter with a Viterbi algorithm, providing an MLSE equalizer for mitigating the nonlinear impairments. Section \ref{results_section} is devoted to numerical results, whereas Sec. \ref{conclusions} is devoted to a discussion and conclusions.

\section{A few words on notation}\label{notation}
Throughout this paper, the following notation is used:
\begin{itemize}
	\item The bra-ket notation, $|x\rangle$, represents 2-element complex column vectors, and is used for dual polarization signals. The Hermitian conjugate of a vector $|x\rangle$ is denoted by $\langle x|$.
	\item Underlined symbols, such as $\underline{x}$, represent column vectors having more than two elements. 
	These do not always have an intuitive physical meaning, and are used merely for mathematical convenience. 
	\item Matrices are denoted by boldface capital letters, such as $\mathbf{X}$, whereas the identity matrix of size $m\times m$ is written as $\mathbb{I}_m$.
	\item The symbol $\mathbb{E}[x]$ denotes an ensemble expectancy operation over $x$. In the case of a vector, or a matrix, the expectancy is performed for each element separately.
\end{itemize}

\section{Achievable information rate with channel equalization}\label{Achievable_information_section}
The achievable information rate is defined as the mutual information between the channel's input and the receiver's output \textit{using a given receiver}, and is a lower bound on the channel's mutual information\cite{secondini2013achievable}.
 In this section, we investigate the AIR of receivers which use adaptive equalization to mitigate the effects of the NLIN.

Within the time-varying ISI model, the $n$-th sample of the waveform received in the channel of interest (COI) is expressed as \cite{dar2015inter,golani2016modeling},
\begin{align}\label{channel_model}
|s_n\rangle= |a_n\rangle + i\sum_{l=-L}^{L}\textbf{R}_l^{(n)}|a_{n-l}\rangle+|w_n\rangle
,\end{align}
where $|a_n\rangle$ are the dual-polarization transmitted symbols, $\textbf{R}_l^{(n)}$ are $2\times 2$ time-varying ISI matrices whose values are determined by the data transmitted over the ICs, and $|w_n\rangle$ is a Gaussian noise process whose variance is determined by the ASE as well as by all of the nonlinear distortions that are not included in the finite ISI term, such as intra-channel distortions (when no back-propagation is applied), FWM, and the ISI generated by matrices $\textbf{R}_l^{(n)}$ with $|l|>L$.
 Under the assumption that all the nonlinear distortions that are not addressed by the equalizer are independent of the channel data, their treatment as an additive Gaussian noise constitutes a worst-case scenario from an information theoretic standpoint, and hence it does not undermine the validity of our analysis. 
Using the analytical framework defined in \cite{golani2016modeling}, the elements of the ISI matrices are expressed as
\begin{align}\label{ISI_matrix_def}
\textbf{R}_l^{(n)}=\left(\begin{array}{cc}
\kappa_l^{(n)}+r_{l}^{(n)} & p_{l}^{(n)}+i q_{l}^{(n)}\\
p_{l}^{(n)}-i q_{l}^{(n)} & \kappa_l^{(n)}-r_{l}^{(n)}
\end{array}\right)
,\end{align}
where the ISI coefficients (which are the elements of the ISI matrices) $\kappa_l^{(n)}$, $r_{l}^{(n)}$, $p_{l}^{(n)}$ and $q_{l}^{(n)}$ are real-valued wide-sense-stationary stochastic processes, whose mean is zero\footnote{In reality the mean of the coefficient $\kappa_0^{(n)}$ is not zero, but its only implication is a constant phase rotation, which is eliminated by any standard receiver. Therefore we ignore it in what follows.
} and whose correlation properties are known and listed in \cite{golani2016correlations,golani2016modeling}. 
It is mathematically convenient to relate to the distributions of the ISI matrices as jointly Gaussian, a choice whose plausibility has been discussed in \cite{golani2016modeling}.

The goal of the receiver is to estimate the actual values of the ISI coefficients, which are to be used subsequently for undoing (equalizing) the ISI effects. Unlike a generic equalization algorithm (e.g. LMS, or RLS), which does not utilize the knowledge of the ISI parameters' statistics, we will use a Kalman-filter-based approach, which relies on explicit prior knowledge of the statistical properties of the ISI terms and a dynamical model for their behavior. The benefit of the equalization is best represented in terms of the extent to which it increases the AIR of the channel, which can be assessed by using the lower-bound presented in the subsection below.

\subsection{A lower bound on the AIR in the case of a partially-known ISI channel\label{MI_section}}
Since the ISI matrices are not deterministically known, the studied scheme falls into the category of communication through a partially-known ISI channel, the capacity of which has been extensively investigated in \cite{medard2000effect}. Our present derivation follows along very similar lines.
Separating the ISI terms into known and unknown parts, Eq. (\ref{channel_model}) can be written as 
\begin{align}\label{separated_channel_model}
|s_n\rangle= |a_n\rangle + i\sum_{l=-L}^{L}\hat{\mathbf{R}}_l^{(n)}|a_{n-l}\rangle+|v_n\rangle+|w_n\rangle
,\end{align}
where $\hat{\mathbf{R}}_l^{(n)}$ is the current estimation of the ISI matrices and 
\begin{align}
|v_n\rangle=  i\sum_{l=-L}^{L}\Delta\mathbf{R}_l^{(n)}|a_{n-l}\rangle
\end{align}
is the residual ISI, composed of the matrices' estimation errors, $\Delta\mathbf{R}_l^{(n)}$ (so that $\hat{\mathbf{R}}_l^{(n)}+\Delta\mathbf{R}_l^{(n)}=\mathbf{R}_l^{(n)}$). 
As shown the appendix, the covariance matrix of $|v_n\rangle$ is $\Lambda_v= \sigma_v^2\mathbb{I}_2$, where the residual NLIN power per polarization is   
\begin{align} \label{residual_ISI}
\sigma_v^2= P\sum_{l=-L}^{L}(\sigma^2_{\kappa_l}+\sigma^2_{r_{l}} +\sigma^2_{p_{l}} + \sigma^2_{q_{l}})
.\end{align}
Here $P$ is the signal power and the terms inside the summation are the respective variances of estimation errors of $\kappa_l$, $r_{l}$, $p_{l}$ and $q_{l}$, described in Eq. (\ref{ISI_matrix_def}). 
Note that although $|v_n\rangle$ is written as additive noise, its variance is proportional to the COI power.

The receiver acquires a sequence of $N$ data points, $|s_{1}\rangle, \cdots, |s_{N}\rangle$ and attempts to detect their corresponding data symbols, $|a_{1}\rangle ,\cdots, |a_{N}\rangle$. It is assumed that the sequence length is much larger than the channel memory, i.e. $N \gg 2L+1$. Using the chain rule representation \cite{book}, the mutual information between transmitter and receiver normalized by the sequence length, can be written as
\begin{align}\label{MI_def}
I \hspace{-0.1cm}&=\hspace{-0.1cm} \frac{1}{N} \hspace{-0.1cm}\left[h\left(|a_{N}\rangle, \cdots \hspace{-0.05cm}, |a_{1}\rangle\right) \hspace{-0.1cm}-\hspace{-0.1cm} h\left(\hspace{-0.05cm}|a_{N}\rangle, \cdots \hspace{-0.05cm}, |a_{1}\rangle \Big| |s_{N}\rangle, \cdots \hspace{-0.05cm},\hspace{-0.05cm} |s_{1}\rangle \hspace{-0.05cm}\right)\hspace{-0.05cm}\right]
\nonumber\\
&=\frac{1}{N}\hspace{-0.1cm}\sum_{n=1}^{N} \hspace{-0.05cm}h(|a_{n}\rangle) - h\left(|a_{n}\rangle \Big| |a_{n-1}\rangle,\cdots\hspace{-0.05cm}, |a_{1}\rangle,|s_{N}\rangle, \cdots \hspace{-0.05cm}, |s_{1}\rangle\right)
,\end{align}
where $h(x)= -\mathbb{E}\left[\log(f_x(x))\right] $ is the differential entropy of a variable $x$, with $f_x(x)$ being its probability density function. 

The mutual information may be lower bounded by removing the conditioning on previous symbols from the conditional entropy of Eq. \eqref{MI_def}. Under the assumption that the symbols are independent and identically distributed, the per-symbol mutual information lower bound is
\begin{align}
I &\ge h(|a_{n}\rangle) - h\left(|a_{n}\rangle \Big| |s_{N}\rangle, \cdots , |s_{1}\rangle\right) 
.\end{align}
We limit the discussion to receivers that are based on equalization, namely the receiver estimates the ISI matrices, $\hat{\textbf{R}}_{l}^{(n)}$, using all of the acquired samples, and then utilizes these estimates and a smaller set of samples for the purpose of estimating the symbols themselves. As the estimated matrices are a function of the received data points (and not the transmitted symbols themselves), we may write
\begin{align} \label{MI_partially_known_ISI}
\hspace{-1cm} h\left(|a_{n}\rangle \Big| |s_{N}\rangle, \cdots , |s_{1}\rangle\right) \le &\nonumber\\
 &\hspace{-3cm} h\left(|a_n\rangle \Big| |s_{n-L}\rangle, \cdots, |s_{n+L}\rangle, \hat{\textbf{R}}_{-L}^{(n)},\cdots, \hat{\textbf{R}}_{L}^{(n)}\right).
\end{align}
%
%
%
Then, using a method similar to that introduced in \cite{medard2000effect}, we 
 show in the appendix that the following relation holds
\begin{align} \label{conditional_entropy_final}
&h\left(|a_n\rangle \Big| |s_{n-L}\rangle \cdots |s_{n+L}\rangle, \hat{\textbf{R}}_{-L}^{(n)}\cdots \hat{\textbf{R}}_{L}^{(n)}\right) \le \\
&\hspace{1.5cm} \log (P) - \log\left( 1+\frac{P}{\sigma_w^2+\sigma_v^2}\right)+ 2 \log (2\pi e)\nonumber
,\end{align}
were $\sigma_w^2$ is variance associated with each of the two elements of the noise term $|w_n\rangle$. This result is equivalent to stating that the receiver can compensate perfectly for the known part of the ISI by equalizing it, and thus only the ISI estimation error and additive noise affect the detection.
Defining the effective signal to noise ratio (ESNR) as
\begin{align}\label{effective_SNR_def}
\mathrm{ESNR}= \frac{P}{\sigma_w^2+\sigma_v^2}
,\end{align} 
and using Eqs. (\ref{MI_partially_known_ISI}) and (\ref{conditional_entropy_final}), we find that the AIR is lower-bounded by
\begin{align} \label{MI_lower_bound}
I \ge \log\left(1+ \mathrm{ESNR}\right) - \Delta h,\end{align}
where the first term on the right-hand-side is the capacity of a Gaussian channel operating with a signal to noise ratio that is equal to the ESNR, and 
\begin{align} 
\Delta h = \left[\log(P)
+ 2\log\left(2\pi e\right) \right]-h\left(|a_n\rangle\right)
,\end{align}
which is the difference between the entropy of our input constellation and that of Gaussian modulation.

Notice that Eq. (\ref{MI_lower_bound}) is valid independently of the particular procedure that is used in order to estimate the ISI matrices. Yet, the quality of the estimation affects the ESNR through the variance $\sigma_v^2$ of the residual ISI terms $|v_n\rangle$, which are treated as noise. Hence, the better the estimation procedure the higher the ESNR and the higher the AIR lower-bound.  
Equation \eqref{MI_lower_bound} illustrates the relevance of the ESNR as a measure for the ultimate performance of equalizers. We will use this measure in Sec. \ref{results_section} in order to evaluate the Kalman-MLSE equalizer, which is to be described in the subsection that follows.

\subsection{A Kalman filter for the optimal tracking of the ISI coefficients\label{kalman_section}}
The Kalman filter is a tool that allows  optimal adaptive estimation of the ISI matrices in the minimum mean square error sense. As briefly pointed out earlier, its advantage over previously attempted methods (such as RLS and LMS) is in accounting for the actual joint second order statistics of the ISI matrix elements. Our construction of the Kalman filter is an adaptation of the method reported in \cite{komninakis2002multi} in the context of the Rayleigh fading wireless channel. 

We define the channel state vector
\begin{align}
\underline{x}_{n}= \left(\kappa_{-L}^{(n)},r_{-L}^{(n)}, p_{-L}^{(n)}, q_{-L}^{(n)},\kappa_{-L+1}^{(n)}\cdots q_{L}^{(n)}\right)
,\end{align}
which consists of all the $4(2L+1)$ ISI coefficients at time $n$, that we attempt to estimate. 
In order to account for the time dependence of the ISI coefficients, several past values of the vector $\underline{x}_{n}$ must be stored. Assuming that the number of relevant past vectors can be limited to $M$, the Kalman procedure requires the consideration of a concatenated state vector,
\begin{align}
\underline{\xi}_n= \left(\underline{x}_n,\underline{x}_{n-1}\cdots \underline{x}_{n-M+1}\right)^T
,\end{align}
whose dimension is $4M(2L+1)\times 1$.
In addition, the state's dynamics needs to be described by a transition equation of the form
\begin{align}
\underline{\xi}_n= \mathbf{F}\underline{\xi}_{n-1} + \underline{z}_n
.\end{align}
where $\underline{z}_n$ is a random Gaussian vector (independent and identically distributed for different values of $n$). The first $4(2L+1)$ elements of  $\underline{z}_n$ represent the random variations of the ISI coefficients, while all other elements are always zero, as they correspond to past values of the coefficients. Proper selection of the matrix $\mathbf{F}$ and the properties of $\underline{z}_n$ will ensure that the second order statistics of the ISI coefficients (the elements of $\underline{\xi}_n$) is consistent with the actual statistics obtained in \cite{golani2016modeling}. 

\begin{figure}[t]
	\centering
	\includegraphics[trim={1cm 12.2cm 4.1cm 0cm},clip,width=9cm]{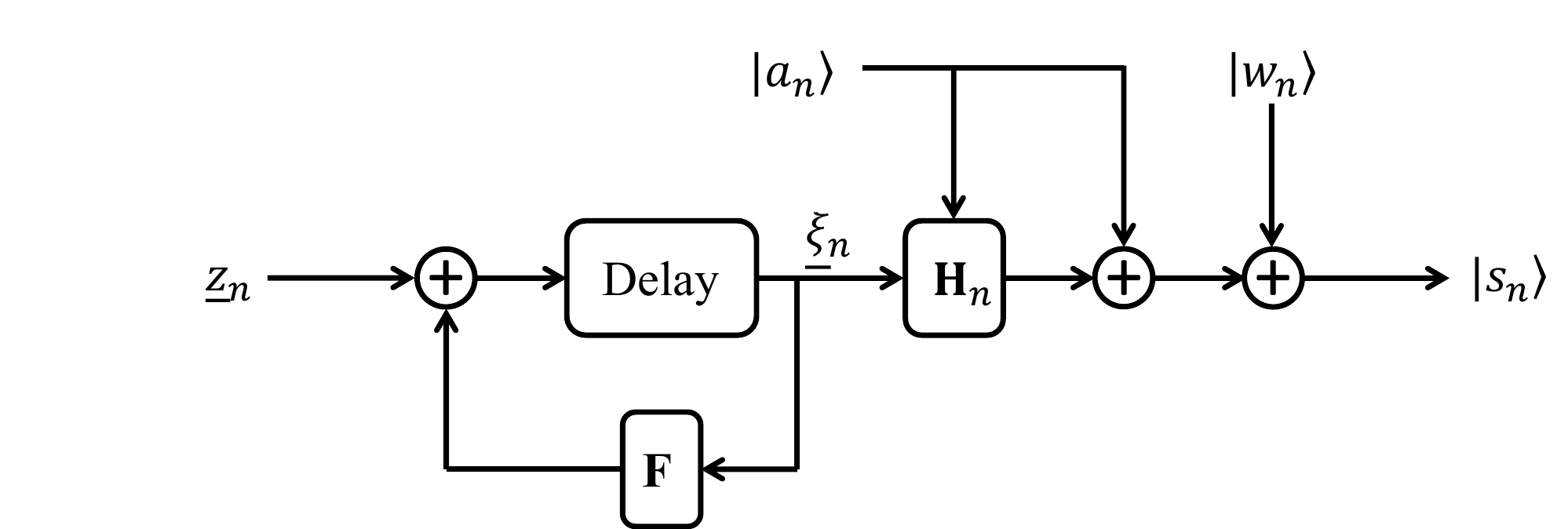}
	\caption{Dynamic system model of NLIN. The ISI coefficients created by the ICs are approximated as autoregressive processes of order $M$, whose output $\underline{\xi}_n$ is the time varying state, to be tracked by the Kalman algorithm.}
	\label{graphical_channel_model}
\end{figure}

Equation (\ref{channel_model}) can be rewritten as
\begin{align}\label{observation_model_Kalman}
|s_n\rangle= |a_n\rangle+ \mathbf{H}_n\underline{\xi}_n+|w_n\rangle
,\end{align}
where $\mathbf{H}_n$ is a $2\times 4M(2L+1)$ matrix that can be written as a concatenation of multiple $2\times4$ matrices as follows,
\begin{align}\label{observation_matrix_def}
\mathbf{H}_n &= i\left(\mathbf{A}_{n-L}, \cdots ,\mathbf{A}_{n+L},0\cdots 0\right)
,\end{align}
where
\begin{align}
\mathbf{A}_m &= \left(\begin{array}{cccc}
a_x^{(m)}& a_x^{(m)} & a_y^{(m)}& -i a_y^{(m)} \\
a_y^{(m)}& -a_y^{(m)} & a_x^{(m)}& i a_x^{(m)}
\end{array}
\right)
,\end{align}
and where $a_x^{(m)}$ and $a_y^{(m)}$ denote the two polarization components of $|a_m\rangle$, and the zeros appended to the symbol matrices are added simply to match the dimensions of the concatenated state vector, $\underline{\xi}_n$. The matrix $\mathbf{H}_n$ is regarded as the time-varying observation model. Together with the state transition matrix $\mathbf{F}$ and the variances of $\underline{z}_n$ and $|w_n\rangle$, it provides all the components that are necessary for a textbook implementation of a Kalman filter \cite{anderson2012optimal}.
Figure \ref{graphical_channel_model} shows a graphical representation of the finite-length model of NLIN.

Lastly, we must find the matrix $\mathbf{F}$ and the variances of $\underline{z}_n$, which will give the correct second-order statistics. We consider the $k$-th element of the state vector $\underline x_n$, which we denote by $x_k^{(n)}$, and whose auto-covariance function by $c_k(m)= \mathbb{E}\left[x_k^{(n)}x_k^{(n+m)}\right]$, can be found in \cite{golani2016modeling}. 
This element can be  approximated  by an auto-regressive (AR) process of order $M$ as follows. Define
\begin{align} \label{single_element_AR}
x_{k,\mathrm{AR}}^{(n)}= \underline{\phi}_k\cdot\left(x_{k,\mathrm{AR}}^{(n-1)}\cdots x_{k,\mathrm{AR}}^{(n-M)}\right)+ z^{(n)}_k
,\end{align}
where $z_k^{(n)}$ is is the $k$-th element of $\underline{z}_n$ and $\underline{\phi}_k$ is a vector of length $M$. The variance of $z_k^{(n)}$ and the values of $\underline{\phi}_k$ are found using 
\begin{align} \label{single_element_AR_paramters}
\sigma_k^2&= c_k(0)-\left(c_k(1)\cdots c_k(p)\right)\mathbf{C}^{-1} \left(c_k(1)\cdots c_k(M)\right)^T \\
\underline{\phi}_k&=\left(c_k(1)\cdots c_k(M)\right)\mathbf{C}^{-1}
,\end{align}
where $\mathbf{C}_{i,j}= c_k(i-j)$ with $0\le i,j<M$, is the covariance matrix of the truncated process.

\begin{figure}[t]
	\centering
	\includegraphics[trim={0cm 9cm 0cm 0cm},clip,width=9cm]{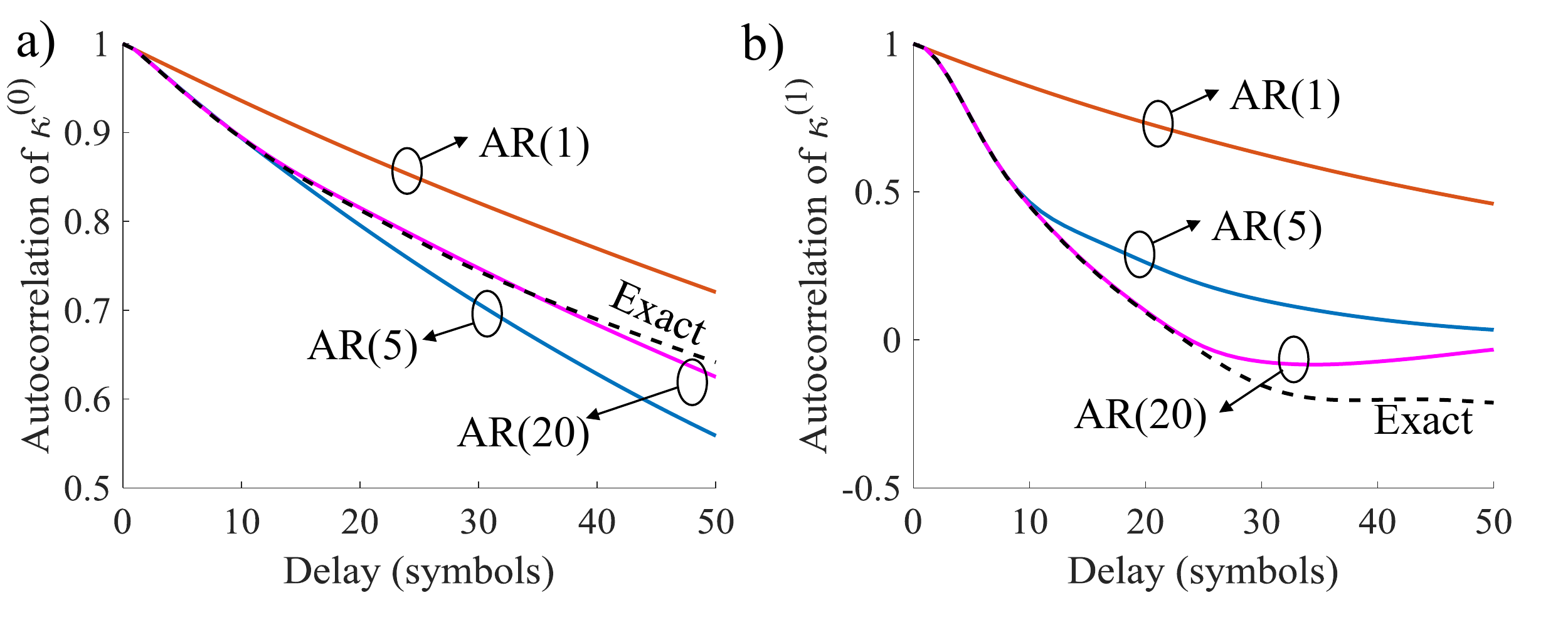}
	\caption{Autoregressive model approximation of NLIN. Showing the autocorrelation function of the first two ISI terms, $\kappa_0^{(n)}$ (a) and $\kappa_1^{(n)}$ (b). Dashed lines show the analytical results obtained using the method described in \cite{golani2016modeling}, solid lines show the autocorrelation functions of the AR(1) and AR(3) approximations.}
	\label{AR_approximation}
\end{figure}

This procedure ensures that $x_{k,\mathrm{AR}}^{(n)}$ has the same auto-covariance function as $x^{(n)}_k$ (up to the memory length $M$), making its behavior a good approximation for the actual NLIN dynamics. This process is repeated for each of the coefficients that we want to track. 
Using the vectors $\underline{\phi}_k$, the state transition matrix, $\mathbf{F}$, is given by
\begin{align}
\mathbf{F}= \left(\begin{array}{ccc}
\mathbf{\Phi}(1)& \cdots& \mathbf{\Phi}(M) \\
\mathbb{I}_{4(M-1)(2L+1)} & & 0\\
\end{array}
\right)
,\end{align}
where $\mathbf{\Phi}(m)$ are a set of $4(2L+1)\times4(2L+1)$ diagonal matrices, satisfying $\mathbf{\Phi}_{i,j}(m)= \phi_i(m)\delta_{i,j}$ (i.e. the $i$-th element on the diagonal is the $m$-th element of $\phi_i$, defined in Eq. (\ref{single_element_AR_paramters})).
Figure \ref{AR_approximation} shows the autocorrelation function of an AR process created this way, for the first two ISI coefficients ($\kappa_0^{(n)}$ and $\kappa_1^{(n)}$), for the example case of 10x100 km link with 80 WDM channels. It is evident that choosing a sufficiently high AR order is very important in order to accurately represent the true behavior of the ISI coefficients. For example, for the $\kappa^{(n)}_1$ coefficient, an AR(1) model quickly diverges from the actual autocorrelation while an AR(3) representation is considerably more accurate. In general, the larger $M$, the broader the range of values in which the autocorrelation of the AR process approximates the actual autocorrelation accurately. In the simulations performed for this paper we used $M=5$, as the benefit from using higher values was too small to justify the increase in computational complexity.
%

%

\section{MLSE receiver\label{MLSE_section}}

An ISI channel with known coefficients can be optimally equalized using the Viterbi algorithm, which constitutes the MLSE of the channel. However, when the coefficients are unknown and need to be estimated from the received signal itself, no efficient rigorous MLSE algorithm exists, and one must resort to the use of approximations \cite{schniter2011equalization,raheli1991principle,chen2003mlse}. 
We opt to implement the MLSE equalizer using per-survivor processing (PSP) Viterbi algorithm \cite{raheli1991principle}, as it suites the sequential nature of the Kalman filter described in the previous section. Simply put, the algorithm maintains a set of Kalman filters, one for each trellis branch, and uses their output to determine the likelihood metric. 

Let the sequence of symbols, estimated by one of the trellis branches, be $|\hat{a}_0\rangle \cdots |\hat{a}_n\rangle$. These are input to the branch's Kalman filter, replacing the actual transmitted symbols in Eq. (\ref{observation_matrix_def}). The filter produces an estimation of the current channel state $\hat{\underline{\xi}}_n$, with an estimation error given by
\begin{align}
|e_n\rangle= |s_n\rangle-|\hat{a}_n\rangle-\hat{\mathbf{H}}_n \hat{\underline{\xi}}_n
.\end{align}
Under the assumption that the correct (error free) symbol sequence was used to create $\hat{\mathbf{H}}_n$, the estimation error is a Gaussian process, whose log-likelihood is
\begin{align}\label{likelihood_metric}
\mathcal{L}_n= -\langle e_n|\Lambda_e^{-1} |e_n\rangle - \log \Lambda_e
,\end{align}
where $\Lambda_e$ is the error's covariance matrix. The branch metric is 
\begin{align}\label{branch_metric}
\mathcal{M}_n= \mathcal{M}_{n-1}+\mathcal{L}_n
,\end{align}
and it is the quantity that the VA attempts to maximize.
An important feature of the Kalman filter is that it inherently calculates both $|e_n\rangle$ and $\Lambda_e$, and so finding the likelihood metric adds very little to the algorithm's computational cost. Equations (\ref{likelihood_metric}) and (\ref{branch_metric}) are accurate only if the estimated symbol sequence is indeed error free. The inherent nonlinearity of hard symbol decision skews the statistics of $|e_n\rangle$, making the errors both mutually dependent and non-Gaussian. Nevertheless, in the high SNR case the influence of symbol errors is small and Eqs. (\ref{likelihood_metric}) and (\ref{branch_metric}) can be used as is.

\section{Numerical results}\label{results_section}

The algorithm has been tested on a variety of links,
 focusing on the case of 80 WDM channels (a setting to which we refer in what follows  as a fully loaded system), carrying a polarization multiplexed 16-QAM modulated signal, operating at a baud-rate of 32 Gbaud with root-raised-cosine pulses having a roll-off coefficient of 0.2, and with channel spacing of 50 GHz. The links  consisted of multiple 100km SMF spans ($D$=17 ps/nm/km, $\gamma$= 1.3/W/km, $\alpha$= 0.2dB/km), separated by lumped amplifiers characterized by a noise figure of 5dB. An additional 3dB of lumped loss was added to each span, so as to account for excess link losses and bring the span loss budget to a reasonable value of  23dB. In all cases, the Kalman filter used a correlation memory of $M=5$ symbols. Links with a number of spans between one and 20 were investigated, so as to produce insight for both metro and long haul scenarios. The simulations performed for  fully loaded systems were done using the virtual lab tool, which was introduced and validated in \cite{golani2016modeling}. We also perform a set of split-step simulations with a smaller number of WDM channels in order to further demonstrate the virtual lab's accuracy.
 
  \begin{figure}[t]
  	\centering
  	\includegraphics[trim={0cm 8.8cm 0cm 0cm},clip,width=9cm]{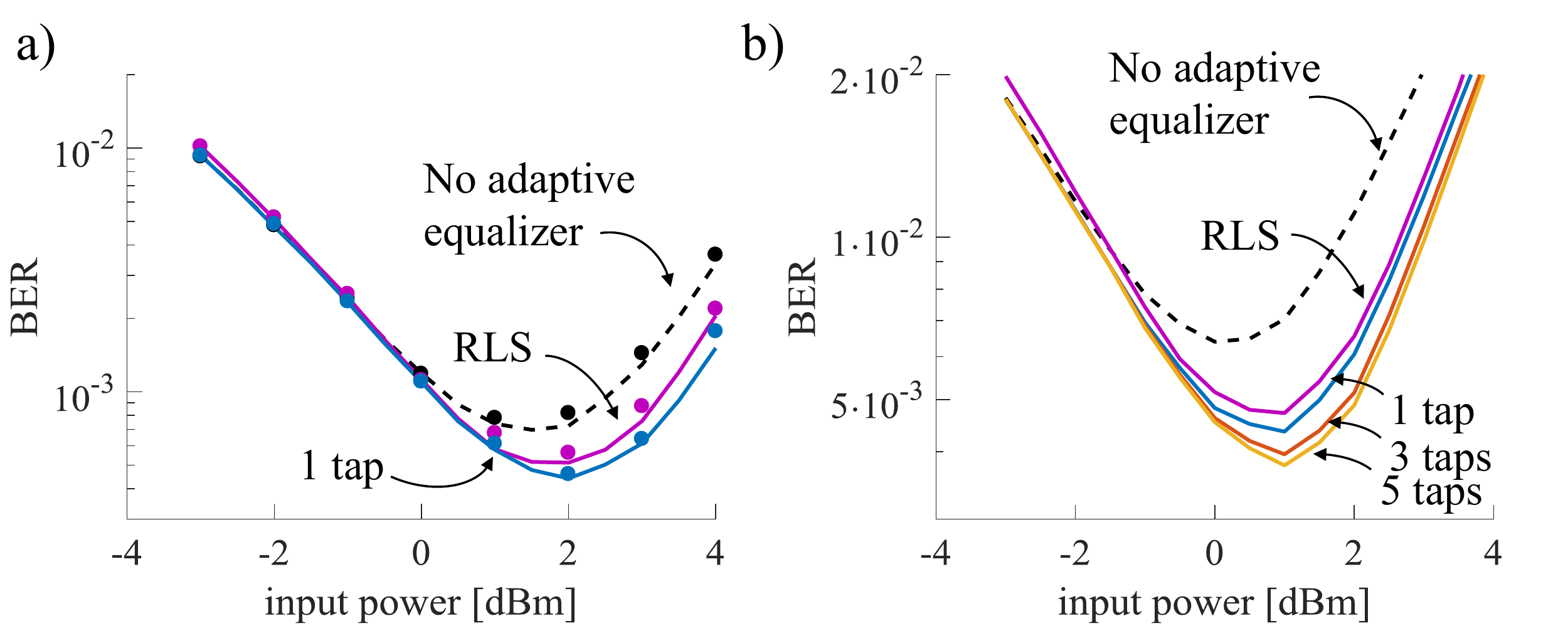}
  	\caption{Comparison of pre-FEC BER performance with different equalizers. (a) results for a $5\times100$km link, carrying 11 WDM channels. (b) results for a fully loaded system (80 WDM channels), over a $10\times100$km link. The receivers employ either RLS equalization (purple), a single tap Kalman-MLSE (blue), or a 3-tap Kalman-MLSE (orange). Results for the simplified receiver which does not compensate for nonlinearity, are shown by the black dashed curve. In (a) Simulations performed using the virtual lab method are shown as solid curves, and split-step simulations are represented by circles.}
  	\label{BER_figure}
  \end{figure}
  
 Figure \ref{BER_figure} shows the BER as a function of the average power per WDM channel, using several receiver types; a simplified receiver which does not employ any nonlinearity equalization, a receiver that uses single tap RLS equalization \cite{dar2015inter}, and receivers employing a 1, 3, and 5 tap Kalman-MLSE equalization. The forget factor for the RLS equalizer was 0.985, which was found to provide the optimal results. Higher order RLS results are not shown, as they provided no further performance improvement. Additionally, since the complexity of the Viterbi algorithm scales exponentially in the channel memory length, using more than 5 taps in the Kalman-MLSE receiver was computationally infeasible.
 
Figure \ref{BER_figure}a aims at validating the virtual lab procedure, and hence it shows the case of 11 WDM channels in a $5\times100$km link, for which split-step simulations are still doable. The circles represent split-step results, whereas the solid curves were obtained using the virtual lab. The excellent accuracy of the virtual lab results is self evident. 
The BER curves of Fig. \ref{BER_figure}b were obtained from the virtual lab, for the case of a fully loaded WDM system over a $10\times100$km link. 
The single tap Kalman MLSE equalizer outperforms the RLS equalizer by 0.1dB in terms of Q-factor\footnote{We use the following definition of Q-factor: $Q= \sqrt{2}\mathrm{erfc}^{-1}(2\cdot \mathrm{BER})$, where $\mathrm{erfc}^{-1}$ is the inverse of the complementary error function.}. 
The additional benefits of 3 and 5 taps in the case of the Kalman MLSE receiver were 0.1dB and 0.06dB, respectively, accumulating to a total gain of 0.6dB compared to the unequalized case. 

\begin{figure}[t]
	\centering
	\includegraphics[trim={0cm 11cm 11cm 0cm},clip,width=6.5cm]{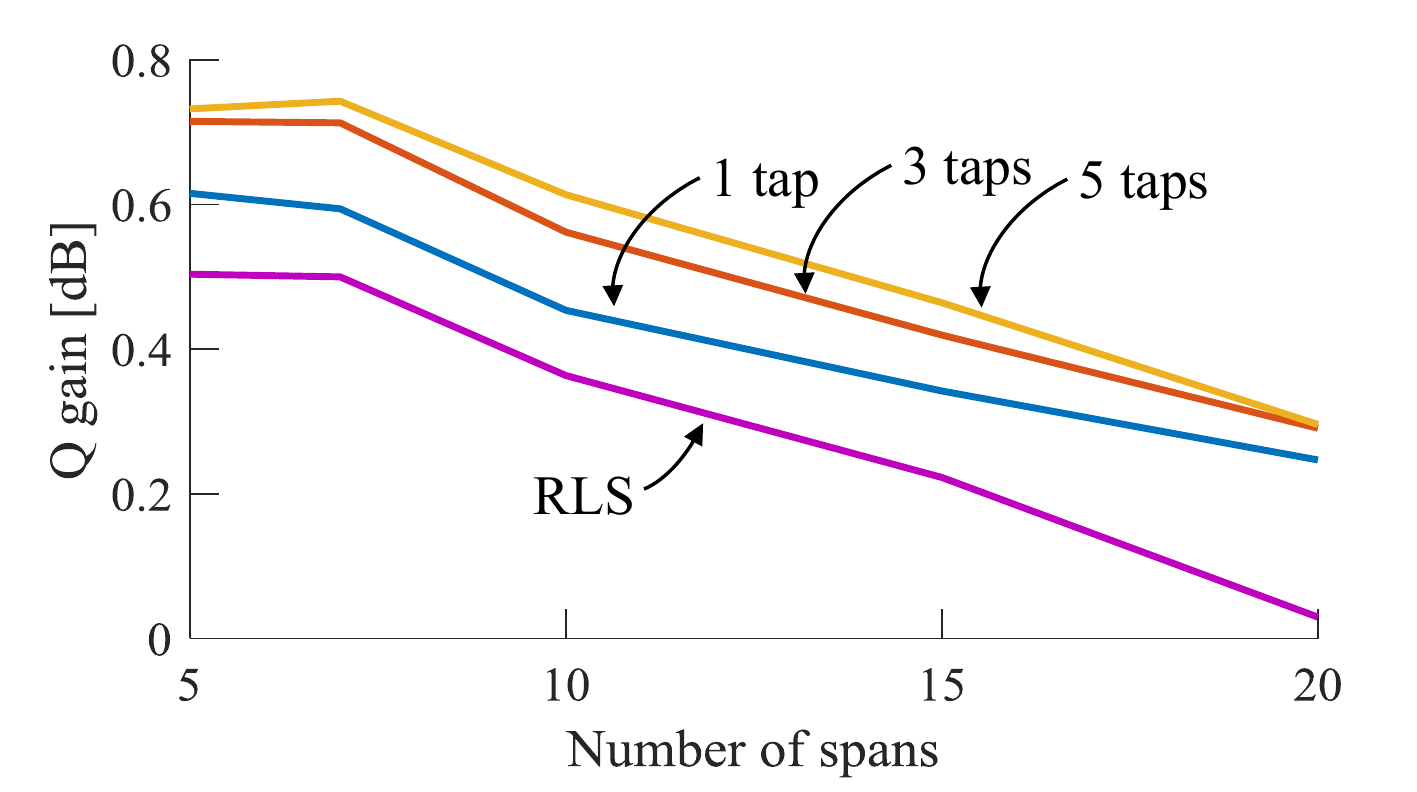}
	\caption{Performance gain for various receivers and link lengths, relative to the case of non nonlinearity compensation. The gain is defined in terms of Q-factor, $Q_{\mathrm{gain}}= {Q_{\mathrm{equ}}}-{Q_{\mathrm{ref}}}$, where all quantities are in decibels. Results are shown for a single tap RLS equalizer and for a Kalman equalizer containing 1,3, and 5 taps.
	}\label{Q_vs_Nspan_figure}
\end{figure}
Figure \ref{Q_vs_Nspan_figure} shows the gain, in terms of Q-factor, relative to the performance of the simplified receiver which does not mitigate NLIN, versus the number of spans. The channel power in all cases corresponds to that providing the minimal BER value. Gains of up to 0.8dB are obtained with the short links, but they reduce as the link length increases. The reason for this reduction is twofold; Firstly, the channel memory becomes longer, and the contribution of the ISI coefficients that we do not compensate for ($|l|>L$) increases. Secondly, the coefficient estimation becomes more difficult due to the higher ASE noise level.

The pre-FEC BER is a good performance estimate in the case of hard-decision receivers. In a more general setting, the proper predictor for the potential performance of a system is the per-symbol mutual information, as pointed out in \cite{alvarado2016replacing}, and discussed in Sec. \ref{MI_section}. As stated in Eq. (\ref{MI_lower_bound}), the mutual information lower-bound is determined by the ESNR, which is bounded in turn by the accuracy with which the equalizer estimates the ISI coefficients. The remainder of this section is devoted to these considerations. 

\begin{figure}[t]
	\centering
	\includegraphics[trim={0cm 8.8cm 6cm 0cm},clip,width=7.2cm]{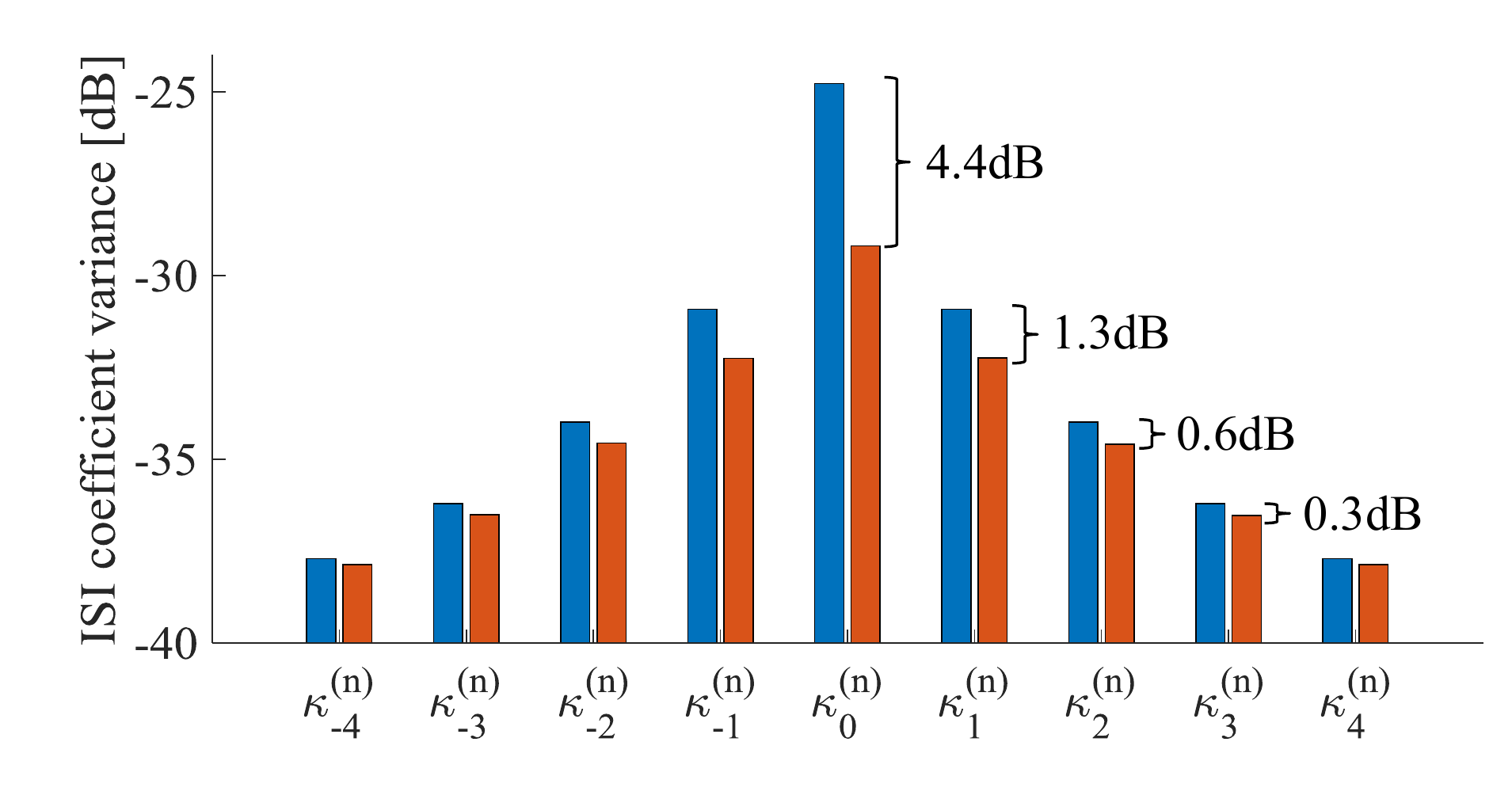}
	\caption{The effectiveness of ISI equalization. The blue and red bars show the mean-square values of the  ISI coefficients $\kappa_l^{(n)}$ before and after equalization. Other elements of the ISI matrix displayed similar behavior.}
	\label{coefficient_tracking_figure}
\end{figure}

In order to assess the accuracy of the estimation of the ISI coefficients, we show in Fig. \ref{coefficient_tracking_figure}  the mean-square values of the coefficients $\kappa_l$ for ($l=-4,\dots,4$) before and after equalization with a 9-tap Kalman filter (tracking matrices $\textbf{R}_{-4} \cdots \textbf{R}_{4}$). The figure was plotted for the case of a $10\times 100$ km link, with an input power of 1dBm. The other elements of the ISI matrices showed qualitatively similar behavior. 

Note that the zeroth order coefficient experiences the largest improvement as a result of equalization, which is of the order of 4 dB. The coefficients $\kappa_{\pm1}$ reduce by approximately 1.3dB, and the effectiveness of equalization continues to deteriorate monotonically for higher ISI orders. It should be noted that the exact numbers appearing in Fig. \ref{coefficient_tracking_figure}, depend on the system parameters, and in particular the effectiveness of equalization increases with launched signal power.

We turn to evaluating the improvement in ESNR in the various settings. Consistently with its definition in Eq. (\ref{effective_SNR_def}), the ESNR at the equalizer's output is given by
\begin{align}
\mathrm{ESNR}= \frac{\mathbb{E}[\langle a_n|a_n\rangle]}{\mathbb{E}[(\langle s_n^{\prime} |- \langle a_n|)(|s_n^{\prime}\rangle - |a_n\rangle)]}
,\end{align} 
where $|s_n^{\prime}\rangle$ denotes the processed samples, which are given by
\begin{align} \label{addedlabel}
|s_n^{\prime}\rangle= |s_n\rangle - i\sum_{-L}^{L}\hat{\textbf{R}}_l^{(n)}|\hat{a}_{n-l}\rangle
,\end{align}
in the presence of NLIN equalization, and $|s_n^{\prime}\rangle= |s_n\rangle$ in its absence. The terms $\hat{\textbf{R}}_l^{(n)}$ and $|\hat{a}_{n}\rangle$ in Eq. \eqref{addedlabel} are the estimated ISI matrices and data symbols respectively.

We evaluate the ESNR in three different settings;
in the first setting, to which we refer as the \emph{simplified scheme}, the symbol estimates $|\hat{a}_{n}\rangle$ in Eq. \eqref{addedlabel} were obtained by applying  hard decisions to the received signal samples, without equalizing the NLIN. In the second setting, to which refer as the \emph{idealized scheme}, the values of $|\hat{a}_{n}\rangle$ are artificially set to the correct symbol values, $|a_{n}\rangle$.
In both the first and the second settings, the symbol estimates are used to construct the Kalman filter, as described in Sec. \ref{kalman_section}. 
In the third setting, the symbols and the ISI matrices are jointly estimated using the Kalman-MLSE procedure, described in Sec. \ref{MLSE_section}. Due to the high computational cost of the Viterbi algorithm, the Kalman-MLSE utilized only three ISI matrices ($\textbf{R}_{0}$, $\textbf{R}_{1}$, and $\textbf{R}_{-1}$) for estimating the data symbols. Then, after the symbol estimates $|\hat a_n\rangle$ were established, additional ISI matrices ($l=\pm2,\pm3,\pm4$) were estimated and used in Eq. \eqref{addedlabel}.  The simplified setting represents a lower bound for what can be obtained with a very reasonable computational complexity, whereas the idealized setting can be viewed as an informal upper bound to the performance of equalization. Yet from a contemporary practical standpoint, the results of the Kalman MLSE form a more relevant upper bound for equalizers with realistic complexity constraints.

\begin{figure}[t]
	\centering
	\includegraphics[trim={0cm 8.8cm 0cm 0cm},clip,width=9cm]{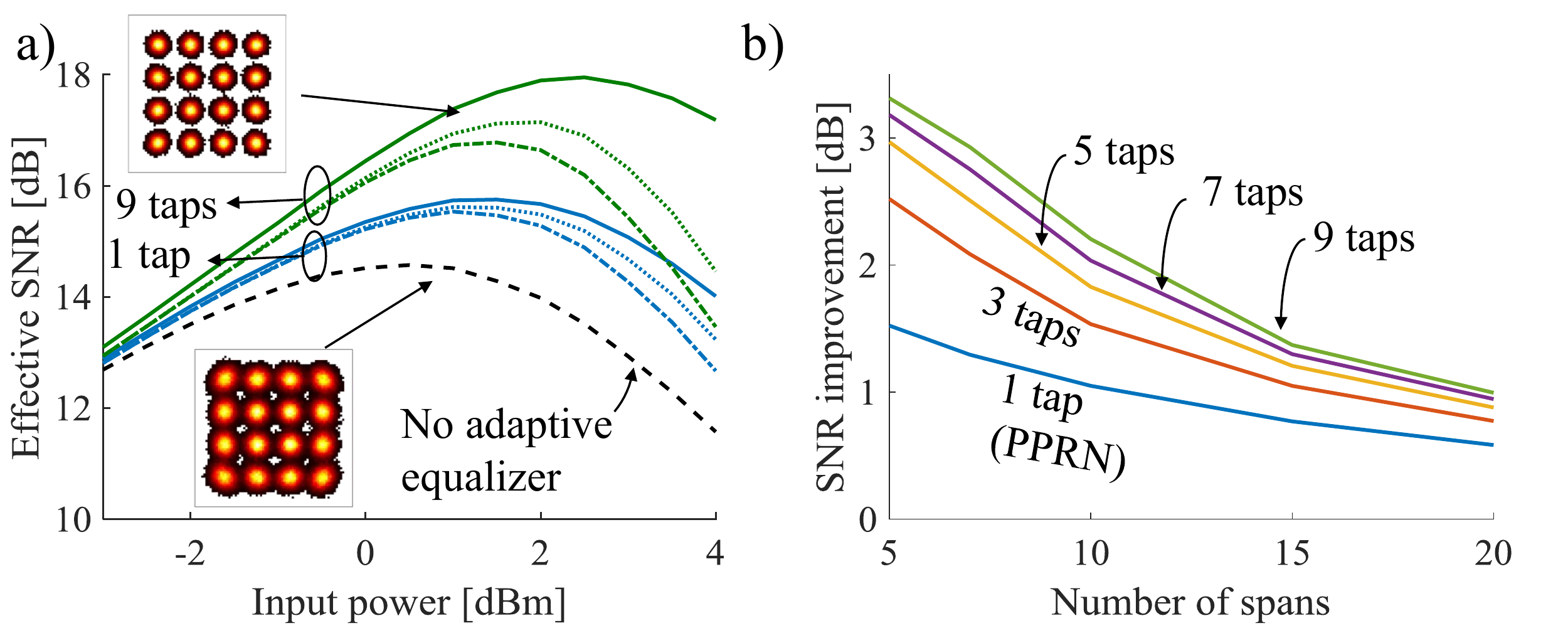}
	\caption{Effective SNR gains of using Kalman filtering. (a) ESNR vs. power, for a $10\times100$km link. The thin solid curve shows the unequalized ESNR. The solid dashed and dotted curves correspond to the the idealized, simplified and regular Kalman MLSE settings, respectively. The insets show output constellation in the unequalized and in teh case oof the 9-tap Kalman-MLSE, both at an input power of 1dBm. (b) The peak ESNR gain versus  number of spans with the Kalman MLSE equalizer.}
	\label{SNR_figure}
\end{figure}

Figure \ref{SNR_figure}a shows the ESNR for a $10\times100$km link, where the dashed black curve shows the unequalized case, which serves as a reference. The blue and green curves correspond to results with a single-tap filter and a 9-tap filter, respectively. The dash-dotted, solid, and dotted curves represent the results obtained with simplified, idealized, and Kalman-MLSE schemes, respectively.
In the case of the single tap equalizer, which compensates solely for PPRN, the simplified and Kalman-MLSE schemes are only within 0.2dB and 0.1dB bellow the ESNR of the idealized scheme, whereas the advantage of the Kalman-MLSE over the unequalized case is of the order of 1.2dB. 
The estimation of higher order coefficients is more sensitive to symbol errors, and for a 9-tap equalizer the differences between the three settings increase. The ESNR of the Kalman-MLSE method is 2.6dB higher than that of the unequalized case, while those of the simplified and idealized settings are 2.2dB and 3.3dB, respectively.

Figure \ref{SNR_figure}b shows the benefit of equalization for different link lengths, using the Kalman-MLSE setting. Gains of more than 3dB are achieved for short links ($500$km), where NLIN is dominated by PPRN. In longer links the relative contribution of PPRN diminishes, and higher order correction becomes more significant. Gains of more than 1dB can be achieved, even for a 2000km link.
The reasons for the overall reduction in equalization gains at longer links are identical to those discussed in the context of Fig. \ref{Q_vs_Nspan_figure}.

\section{Discussion and conclusions\label{conclusions}}
We studied the potential benefit of equalization for the purpose of mitigating inter-channel nonlinear distortions in fiber-optic WDM systems. The benefits of equalization follows from the fact that inter-channel nonlinearities can be modeled as a time-varying ISI process. Yet, the rapid variation of the ISI coefficients limits the efficiency of generic ISI equalization methods. In order to assess the limits of the performance of equalization we have considered an equalizer that is based on the combination of Kalman filtering and MLSE, and which takes advantage of the knowledge of the particular second-order statistics of the ISI matrices characterizing the channel. 

We used two metrics for assessing the efficiency of equalization, one is the pre-FEC BER (or equivalently -- the Q-factor), and the other was the effect of equalization on the ESNR of the channel. The first metric is appropriate for systems using hard decision FEC, but it is less meaningful in systems in which soft-decision FEC is deployed. The second metric is of higher relevance from a fundamental standpoint because it is directly linked to the AIR of the system. In the tested scenarios, the use of the Kalman MLSE receiver improved the Q-factor by up to 0.8dB, whereas the observed ESNR improvement was up to 3.3 dB.   

An important aspect of the considered approach is the complexity of the equalization method. The complexity of the considered Kalman-MLSE equalizer is much higher than what is customary in real-time applications today. For this reason, the performance of the Kalman-MLSE receiver can be considered as a practical upper bound for the performance of equalization. 
It should be noted that most of the computational complexity is associated with the implementation of the Viterbi MLSE algorithm, and not with the Kalman estimation process itself. In view of this, less complex equalizers, that rely on Kalman estimation of the ISI matrices, can also be implemented. The simplified Kalman receiver discussed in the context of Fig. \ref{SNR_figure} exemplifies this approach. 


\appendices
\setcounter{section}{-1} 
\section{Derivation of conditional entropy upper bound \label{Entropy_appendix}}
We will now derive the upper bound on the conditional entropy used in Eqs. (\ref{residual_ISI},\ref{conditional_entropy_final}), based on the procedure described in \cite{medard2000effect}. We start by finding the covariance matrix of the residual ISI term $|v_n\rangle$ from Eq. (\ref{separated_channel_model}). Writing it explicitly with the different polarization components,
\begin{align}
&|v_n\rangle = 
\sum_{l=-L}^{L} |v_{n-l}\rangle = \\
 & i\sum_{l=-L}^{L} \hspace{-0.2cm}\left(\begin{array}{c}
\hspace{-0.3cm} \left(\Delta\kappa_l^{(n)}+\Delta r_{l}^{(n)}\right) a_{n-l,x} + 
\left(\Delta p_{l}^{(n)}+i \Delta q_{l}^{(n)}\right)  a_{n-l,y} \\
\hspace{-0.2cm}\left(\Delta p_{l}^{(n)}-i \Delta q_{l}^{(n)} \right) a_{n-l,x} + 
\left(\Delta \kappa_l^{(n)}-\Delta r_{l}^{(n)}\right) a_{n-l,y} 
\end{array}\hspace{-0.2cm}\right) \nonumber
.\end{align}
Looking at each element of the summation separately and using the fact that  the data symbols in both polarizations are independent and that the coefficients $\kappa_l^{(n)}$, $r_{l}^{(n)}$, $p_{l}^{(n)}$ and $q_{l}^{(n)}$ are uncorrelated, we can write the expectations
\begin{align}
\mathbb{E}\left[ |v_{n-l,x}|^2\right] &= \left(\mathbb{E}\left[ |\Delta \kappa_l^{(n)}|^2\right]+\mathbb{E}\left[ |\Delta r_{l}^{(n)}|^2\right]\right)\mathbb{E}\left[ |a_{n-l,x}|^2\right] \nonumber\\ 
& +
\left(\mathbb{E}\left[ |\Delta p_{l}^{(n)}|^2\right]+\mathbb{E}\left[ |\Delta q_{l}^{(n)}|^2\right]\right)\mathbb{E}\left[ |a_{n-l,y}|^2\right] \nonumber\\ 
&= P(\sigma^2_{\kappa_l}+\sigma^2_{r_{l}} +\sigma^2_{p_{l}} + \sigma^2_{q_{l}}) \\
 & \hspace{-1.8cm} \mathbb{E}\left[ v_{n-l,x}v_{n-l,y}^*\right] = 0
,\end{align}
As the data symbols at different time slots are also independent, the covariance matrix of $|v_n\rangle$ is obtained by summing those of $|v_{n-l}\rangle$, namely 
\begin{align}
\Lambda_v= \sum_{l=-L}^{L} P(\sigma^2_{\kappa_l}+\sigma^2_{r_{l}} +\sigma^2_{p_{l}} + \sigma^2_{q_{l}})\mathbb{I}_2= \sigma_v^2\mathbb{I}_2
.\end{align}

Next, we turn to evaluate the conditional entropy in Eq. (\ref{MI_partially_known_ISI}),
\begin{align}
	 h_c= h\left(|a_n\rangle \Big| |s_{n-L}\rangle \cdots |s_{n+L}\rangle, \hat{\textbf{R}}_{-L}^{(n)}\cdots \hat{\textbf{R}}_{L}^{(n)}\right)
.\end{align} 
Let $|\hat{a}_n\rangle$ be an estimator of $|a_n\rangle$ which uses only the condition variables $|s_{n-L}\rangle \cdots |s_{n+L}\rangle, \hat{\textbf{R}}_{-L}^{(n)}\cdots \hat{\textbf{R}}_{L}^{(n)}$. As $|\hat{a}_n\rangle$ is a function of only the condition variables, it follows that
\begin{align}
	h\left(|a_n\rangle \Big| |s_{n-L}\rangle \cdots |s_{n+L}\rangle, \hat{\textbf{R}}_{-L}^{(n)}\cdots \hat{\textbf{R}}_{L}^{(n)}\right) \le 	h\left(|a_n\rangle \Big| |\hat{a}_n\rangle\right)
.\end{align} 
The entropy of any continuous random variable may be bounded by the entropy of a Gaussian variable with the same variance, and so  
\begin{align}\label{Gaussian_bound}
h_c \le h\left(|a_n\rangle \Big| |\hat{a}_n\rangle\right) \le \frac{1}{2} \log |\Lambda_{z}| + 2 \log (2\pi e)
,\end{align}
where $\Lambda_{z}$ is the $2\times 2$ conditional covariance matrix of $|a_n\rangle$ given $|\hat{a}_n\rangle$. 
Consider a suboptimal detector employing a zero-forcing equalizer, 
\begin{align}\label{zero_forcing_equ}
|\hat{a}_n\rangle &= |s_{n}\rangle - i\sum_{l=-L}^{L}\hat{\mathbf{R}}_l^{(n)}|s_{n-l}\rangle \\
&= |a_{n}\rangle + |w_{n}\rangle + |v_{n}\rangle -i \sum_{l=-L}^{L}\hat{\mathbf{R}}_l^{(n)}|N_{n-l}\rangle\nonumber
.\end{align}
were $|N_{n}\rangle= |s_{n}\rangle-|a_{n}\rangle$ is the noise terms of sample $|s_{n}\rangle$, including both NLIN and ASE. The summation in Eq. (\ref{zero_forcing_equ}) corresponds to the equalizer's noise enhancement term, which is given by 
\begin{align}
|N_e\rangle &= \sum_{l=-L}^{L}\hat{\mathbf{R}}_l^{(n)}|N_{n-l}\rangle= \sum_{l=-L}^{L}\hat{\mathbf{R}}_l^{(n)}\left(|w_{n-l}\rangle + |v_{n-l}\rangle\right) \nonumber\\
&+ \sum_{l,l'=-L}^{L}\hat{\mathbf{R}}_l^{(n)}\hat{\mathbf{R}}_{l'}^{(n-l)}|a_{n-l-l'}\rangle
.\end{align}
The noise enhancement' covariance matrix is 
\begin{align}\label{noise_enhancement_variance}
\mathbb{E}\left[|N_e\rangle \langle N_e|\right]&= (\sigma_w^2+\sigma_v^2) \sum_{l=-L}^{L}\hat{\mathbf{R}}_l^{(n)}\hat{\mathbf{R}}_l^{(n)\dagger} \\
&+ P \sum_{l,l'=-L}^{L}\hat{\mathbf{R}}_l^{(n)}\hat{\mathbf{R}}_{l'}^{(n-l)}\hat{\mathbf{R}}_{l'}^{(n-l)\dagger}\hat{\mathbf{R}}_l^{(n)\dagger} \nonumber
.\end{align}
Recall that the ISI model is only accurate in the framework of first order perturbation theory, and thus the NLIN interference terms must satisfy 
\begin{align}
\left|\sum_{l=-L}^{L}\hat{\mathbf{R}}_l^{(n)}\hat{\mathbf{R}}_l^{(n)\dagger} \right| \ll 1
.\end{align}
This means that the noise-ISI interaction (first term in Eq. (\ref{noise_enhancement_variance})) is negligible with respect to the noise variances $\sigma_w^2$ and $\sigma_v^2$. Furthermore, the ISI-ISI interaction term (second term in Eq. (\ref{noise_enhancement_variance})) contains only elements of the fourth power in the nonlinear coefficients, and is therefore also negligible under the same conditions. Thus, we can ignore the noise enhancement terms in Eq. (\ref{zero_forcing_equ})  and obtain
\begin{align}\label{zero_forcing_equ_approx}
|\hat{a}_n\rangle &\approx |a_{n}\rangle + |w_{n}\rangle + |v_{n}\rangle 
.\end{align}
With this approximation we can find the conditional variance as
\begin{align}
\Lambda_z= \mathbb{E}\left[|a_n\rangle\langle a_n| \Big| |\hat{a}_n\rangle\right]= \frac{P(\sigma_w^2+\sigma_v^2)}{P+\sigma_w^2+\sigma_v^2}\mathbb{I}_2
,\end{align}
and so the bound in Eq. (\ref{Gaussian_bound}) is
\begin{align}\label{Gaussian_bound_final}
h_c \le \log (P) - \log\left( 1+\frac{P}{\sigma_w^2+\sigma_v^2}\right)+ 2 \log (2\pi e)
.\end{align}
The approximation of neglecting noise enhancement is equivalent to assuming that the equalizer compensates perfectly for the known part of the ISI. The zero-forcing equalizer is not optimal for this purpose, and is used here only because of its mathematical simplicity. Using more sophisticated methods, such as a linear least mean squares and decision feedback equalizers can improve the approximation in Eq. (\ref{zero_forcing_equ_approx}), at the cost of less compact results. 

\section*{Acknowledgment}
The authors are pleased to acknowledge financial support from the Israel Science Foundation Grant 1401/16. 

\ifCLASSOPTIONcaptionsoff
\newpage
\fi

 

\end{document}